\documentclass[sigconf,nonacm]{acmart}

\usepackage{booktabs} 
\usepackage{bookmark}
\usepackage{bookmark}
\usepackage{enumitem}
\usepackage{multirow}
\usepackage{balance}
\usepackage{subfig, float, epsfig, epstopdf, balance, algorithm}
\usepackage{amsmath}

\newcommand{\cB}{\color{black}}

\newcommand{\plus}{\raisebox{.4\height}{\scalebox{.6}{+}}}

\def\P{\mathbb{P}}

\renewcommand\footnotetextcopyrightpermission[1]{} 
\setcopyright{none}

\acmYear{2016}

\begin{document}
\title{Location Aware Embedding for Geotargeting in Sponsored Search Advertising}

\author{Jelena Gligorijevic, Djordje Gligorijevic, Aravindan Raghuveer, Mihajlo Grbovic}
\affiliation{%
 \institution{Work done while authors were with Yahoo! Research.}
}

\author{Zoran Obradovic}
\affiliation{%
 \institution{Temple University}
}
\email{zoran.obradovic@temple.edu}

\begin{abstract}
Web search has become an inevitable part of everyday life. Improving and monetizing web search has been a focus of major Internet players. 
Understanding the context of web search query is an important aspect of this task as it represents unobserved facts that add meaning to an otherwise incomplete query.
The context of a query consists of user's location, local time, search history, behavioral segments, installed apps on their phone and so on. 
Queries that either explicitly use location context (eg: ``\textit{best hotels in New York City}'') or implicitly refer to the user's physical location (e.g. ``\textit{coffee shops near me}'') are becoming increasingly common on mobile devices. 
Understanding and representing the user's interest location and/or physical location is essential for providing a relevant user experience. 
In this study, we developed a simple and powerful neural embedding based framework to represent a user's query and their location in a single low-dimensional space. We show that this representation is able to capture the subtle interactions between the user's query intent and query/physical location, while improving the ad ranking and query-ad relevance scores over other location-unaware approaches and location-aware approaches.
\end{abstract}

%
%

\keywords{Sponsored search advertising; distributed representations; word embeddings; machine learning}

\maketitle

\section{Introduction}
 
There has been tremendous growth of Internet usage and Web environments in the previous decade, bringing immense value in the form of quick and widespread information access. 
Such development of size, accessibility, and subsequently, content allowed major Internet players to monetize through advertisements while improving users' experience on Web search (a model called sponsored search). 
In sponsored search, publishers (Web service providers) control the top results of a search so as to display only the most relevant ads that resonate with query intent. 

Modeling and predicting users' query intent has drawn the attention of many researchers \cite{Aiello2016,djuric2014,grbovic2015sigir,ordentlich2016cikm, yin2016}. Recently, notable success was achieved by the family of neural embeddings approaches \cite{grbovic2016sigir} that draw benefit from context of user queries and interactions between users and services, like organic or sponsored link clicks or views for a users' query.

Location plays a huge role in Web search context approximation and in implementing context-aware applications, especially for mobile users. 
Increased volume of mobile Web search has brought unprecedented opportunities for advertisers to personalize their messages to consumers based on their location, in real time, a practice called \textit{geotargeting}.
Users are more likely to become consumers if they receive targeted message by the advertiser that is in the context of their current location, or location they are interested about. 
Proximity of users to points of interest, such are stadiums, airports, universities, and malls, can provide meaningful context to advertisers who target specific interest groups. For example, users that are located in a stadium could be targeted with sport merchandise, as they are likely to identify with their favorite players who wear such apparel.
An important aspect of geotargeting, in addition to capturing physical location is capturing users' local intent that is commonly included in search queries. Users often use location keywords to narrow down their search which are suitable for targeting, i.e. ``\textit{best hotels in New York City}'' or ``\textit{coffee shops near me}''. 
Exploitation of such local intent can greatly help in satisfying users' information needs, improving their experience, and likely leading to significant performance improvements in sponsored search advertising. 

\begin{figure}[h!]
	\begin{center}
		\includegraphics[width=0.6\columnwidth]{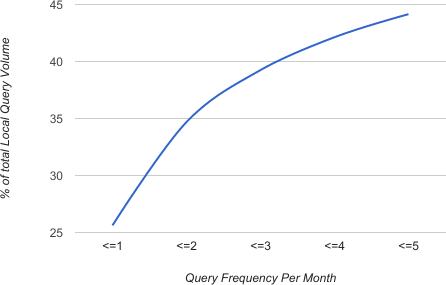}
	\end{center}
	\caption{{\footnotesize Tail is predominant in local intent queries. Queries that occur less than 5 times per month account for close to 45\% of total local query volume}}  
	\label{fig:tail}
\end{figure}

However, modeling local queries has its specific challenges.
Different from other query categories, local queries typically have three components: query intent or query subject, location of interest and implied ranking signals. For instance, the query ``\textit{best hotels in New York}'' implies that the users have intent to inform themselves about hotels (\textit{subject}) in New York (\textit{location}) while taking into account their ratings (\textit{ranking})- best hotels. 
Matching of local queries to advertisements (or any content) requires understanding of these three components of the query.
Another challenge is the heavy tail that is predominant in local intent queries. Local queries that occur just once account for 25\% of the local query volume in a month (Figure~\ref{fig:tail}). Due to a heavy tail, query history based methods~\citep{grbovic2016sigir} cannot address such a query volume. 

Our goal in this study is to model local queries in a way that location information is used to improve understanding of query intent and query components. With improved understanding of query components we aim to solve the problem of heavy tail queries via broad match search. Towards these objectives, we propose a neural embedding family of models, named \textit{worLd2vec} (based on the \textit{search2vec} model \cite{grbovic2016sigir}). The models are using local intent search sessions to learn representations of queries, locations, ads, semantic query fragments and link clicks in a common space, such that they can later be used to retrieve ad vectors for a given local query through a nearest neighbor search on the cosine distance metric.

We make the following contributions in this paper:
\begin{itemize} [noitemsep,topsep=3pt,parsep=3pt,partopsep=0pt]
\item{We propose a unified framework \textit{worLd2vec} for representing Web session context: queries, locations, ads, semantic query fragments and link clicks in a common space, by modeling location information in a global (session) or local (query) context, using five different approaches ($\mathbf{gw2v_{woeid}}$, $\mathbf{gw2v_{poi}}$, $\mathbf{lw2v}$, $\mathbf{lw2v+}s$, $\mathbf{lw2vCRF+}$) introduced in Section \ref{sec:location_web_search}}	
\item In order to address the cold start problem, remove query writing noisiness and to properly model local query components, we propose a composition model $\mathbf{lw2vCRF+}$ from the \textit{worLd2vec} family of models (Section~\ref{sec:lw2v_crf+_sessions})
\item We trained these five location aware search embedding models using  three months of local intent Web search sessions, resulting in ad, location and query representations of very high quality. Extensive evaluation on a following month of real-world search traffic showed that the proposed approaches significantly outperformed the existing state-of-the-art search methods in precision and Normalized Discounted Cumulative Gain (NDCG).
\end{itemize}
In the following section we introduce a problem of query-to-ad matching in sponsored search sessions with local queries. Further, we propose different modeling approaches for understanding Web session context of local queries  in Section~\ref{sec:location_web_search}. We first show solution where location information is used as a global session context, discuss its advantages and shortcomings, and then introduce solutions where location is used as a local query context. We cover related
work in Section~\ref{sec:related+word} and conclude the study in Section~\ref{sec:conclusion}.

\section{Problem set-up}
The main goal of this study is to improve ad ranking and query-ad relevance scores for local intent queries in sponsored search advertising by using information rich location representations while learning neural embeddings of queries, locations, ads, semantic query fragments and link clicks in the same vector space.

\subsection{ System Overview}
\label{sec:system_overview}
\begin{figure}[t!]
	\begin{center}
		\includegraphics[width=0.9\linewidth]{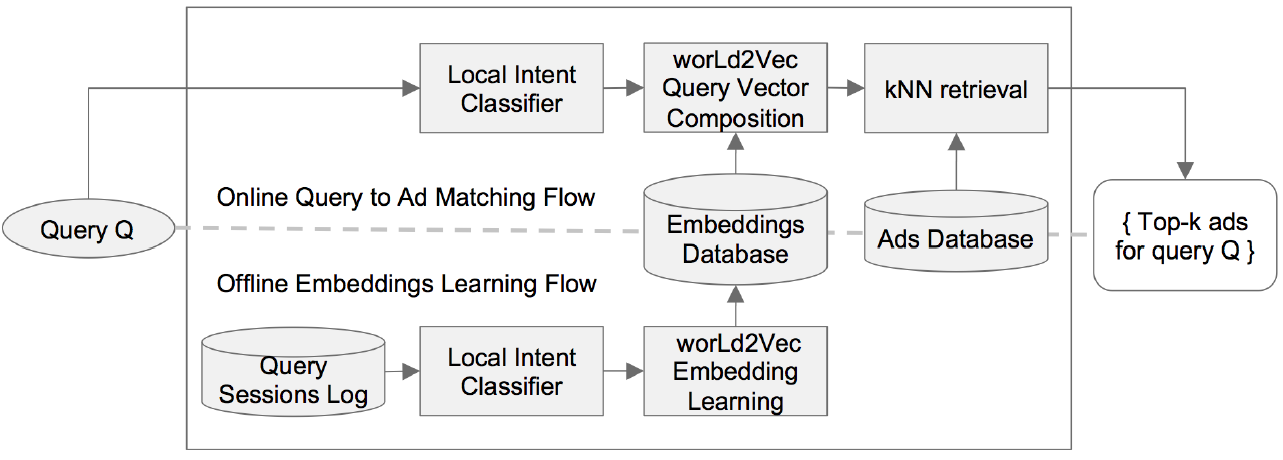}
	\end{center}
	\caption{{\footnotesize {System Architecture:}  Our system consists of two parts: The {\bf\textit{ Offline Embeddings Learning Flow}}  and the  {\bf \textit{Online Query to Ad Matching Flow}}.  
			 In the {\bf \textit{Offline Embeddings Learning Flow}} we use session query and click logs to learn low dimensional embeddings for queries, ads, location and semantic text
			fragments in the same vector space.
			 In the {\bf \textit{Online Query to Ad Matching Flow}} 
			 we use the learned embeddings to retrieve top k ads for an incoming query Q.   The {\em worLd2vec embedding learning} step  in the offline flow and the {\em worLd2vec Query Vector Composition} step in the online flow (both described in Section~\ref{sec:location_web_search}) form the main focus of this paper.  The offline flow is applied only on local queries as tagged by the Local Intent Classifier ~\cite{Yi2009}.
			  The vectors learned by the worLd2vec training step are stored in a embedding database and used later in the  \textit{Online Query to Ad Matching Flow}.  
			  The query vector is at the time of searving used to retrieve the top k nearest ad vectors from the ads database. 
			  }}
	\label{fig:arch} 
\end{figure}
Figure~\ref{fig:arch} shows the overall system architecture.  Our system consists of two parts: \textit{The offline Embeddings Learning Flow} (offline flow, in short) and \textit{the Online Query to Ad Matching flow}(online flow, in short). The focus of this paper is the embeddings learning flow. The online flow is used to evaluate the performance of the offline models.  
The offline flow learns embeddings for local queries, locations, ads, semantic query fragments and link clicks in the same vector space. Detection of local queries is done by a  local intent binary classification model~\cite{Yi2009}. Section~\ref{sec:location_web_search} explains how the embeddings for all proposed approaches are learned in more detail. 
In the online flow, the original query is passed further on to the worLd2vec model to generate a vector representation of the query. 
The vector representation of the query is then used to retrieve ads from the back--end through a nearest neighbor search on the cosine distance metric. 
A detailed discussion of the feature engineering and performance tuning of the Local intent classifier component is out of scope for this paper, however in the following subsection we provide some essential observations.

\subsection{Queries with Local Intent} 
Detecting if a query has local intent (\textit{local query}) has been well studied in the research community~\cite{Venetis2011, Yi2009}.  In the context of local queries we make the following observations:
\begin{enumerate}
\item{A local query can be implicit or explicit. Given the query ``\textit{coffee shops near me}'', the retrieved ads should refer to coffee shops that are near to the user's current location. We refer to such queries as \textit{implicit} local queries. Queries with specific location names (e.g., ``\textit{best hotels in New York City}'') are referred to as \textit{explicit} local queries.}
\item{ A local query can be further categorized into {\em organization queries} and  {\em business category queries}.  Organization queries are those in which the user is looking for a particular business (eg. ``{\em Macy's Mountain View opening hours}''). Business category queries are those queries where the user is looking for a list of organizations belonging to a particular business category (eg. ``{\em coffee shops near me}''). In this case the user wants to see as result a variety of coffee shops (Starbucks, Peet's Coffee, Philz Coffee, Dunkin Donuts) sorted by distance to her physical location.}
\item{Local queries often have a sorting criteria based on which the user wants the results to be ordered. This criteria expressed in terse natural language by the user should be mapped to an attribute of the organizations retrieved and then sorted on that attribute. For instance, the query ``{\em best hotels in New York City}'', implies that the user wants the sorting function to consider the ratings of the hotels in New York city. We term the  query phrases that refer to the sorting criteria (best, cheap, near) as {\em qualifiers}}. 
\item{Some Local queries look for a particular facet of an organization. For instance, the query ``{\em Macys  Mountain View opening hours}'' is looking to find the store hours for a particular organization of interest. The query phrases that refer to organization facets are termed as {\em attributes}}.
\end{enumerate}

\subsection{Location Enriched Sponsored Search Session}
\label{sec:local_queries_sessions}
To train the proposed models, we use search query logs organized in a dataset containing S search sessions. Each session log contains user activities $h$ in the form of search queries, link clicks, or ads ordered by the time of appearance: $s=(h_1,...,h_M)\in S$.

In addition to the search user activities, if the search query had local intent, user location (if allowed) and/or query location $h_{loc}$ is logged, too, based on the nature of the local query. For implicit queries (i.e.``\textit{coffee shops near me}''), we use physical \textit{user} location to learn location embedding, while for explicit local queries (i.e., ``\textit{best hotels in New York City}'')  \textit{query} location is used to learn location embedding for local models. For both types of local queries, location is recorded in the form of zip, city, state, etc. using woeids. 
In addition, we analyze semantic location in a part of our experiments (global models in Section~\ref{sec:lw2v+_sessions}) using \textit{geofencing}, a practice of considering user proximity to a Point of Interest (poi). 

\subsubsection{Physical vs Semantic Location Representation}
Location information can be observed as either physical or semantic. We define \textit{physical location} as the actual location of the user for a given query (i.e. zip or city), while the \textit{semantic location} is defined as proximity to a point of interest (i.e. sports stadium or a mall).

We represent physical location using the Where On Earth IDs (woeids), unique hierarchical identifiers defined over layers, top-to-bottom, as: Earth, continents, countries, states, cities, zip-levels, neighbors and streets.
Locations used for enriching Web sessions in our study are selected to be of the lowest available granularity (up to and including city level).

In addition to the physical location identifier of different granularity found in the woeid, we experimented with users proximity to a Point Of Interest (poi). Pois are defined as specific regions or point locations that someone may find useful or interesting. They can range from places of gathering such are stadiums, malls, airports or places of worship to natural landmarks, such are forests and national parks. 
Spatial polygons for $329,172$ such places across the continental USA are defined. 
For each user's query with assigned latitude and longitude we evaluate whether it was taken in the proximity of a poi using geofencing, where proximity is defined as a walkable distance to the poi, which we estimate as the 25 meters.

\subsection{Ad Retrieval Task Evaluation}\label{sec:eval}

For all of the models, three months of Web search sessions were used for learning embeddings. US only sessions with location (poi or woeid) are kept, while logs with bots, short life cookies and sessions with more than $30$ queries are removed. 
The learned representations of queries were then used to produce list of most relevant ads to a given query. The models were evaluated on local search logs from the testing month in terms of ranking  and  relevance, measured by \textit{precision@K} and \textit{NDCG}, and compared to the state-of-the-art method from the literature.

\paragraph{Measuring Precision@K for Top K Ads }
precision@K is computed for each query as a fraction of relevant (clicked) ads within the $K$ retrieved ones, and average precision is reported.

\paragraph{Measuring Ad Ranking Quality via Editorial Judgments}
Relevance of ads for given queries is analyzed by leveraging professional editors' judgment. Editors manually judge each query--ad pair, assigning one of six grades: Perfectly Relevant, Highly Relevant, Relevant, Somewhat Relevant, Barely Relevant or Irrelevant. 
Once grades for query-ad are obtained, it is possible to provide a gold standard ranking. 
Then Normalized Discounted Cumulative Gain (NDCG) as the metric can be used to evaluate the query to ad matching performance \cite{Aiello2016}.

\section{Location Aware Embeddings for  Sponsored Search Advertising}
\label{sec:location_web_search}

In this section we first describe distributed representation models successfully used on Web search sessions data \cite{grbovic2016sigir} and later we propose five approaches to use location information in addition to a query, search clicks and ads from sponsored search sessions to improve on the local intent query2ad retrieval task.
We introduce them in an iterative manner, where performance with advantages and modeling challenges are discussed for each of the models, leading to development of multiple approaches for different aspects of modeling location context of Web search queries.

\subsection{Distributed Representations for Web Search: s2v} 
\label{sec:search2vec}  
\begin{figure*}[t!]
	\centering
	\subfloat[{\small location agnostic: search2vec}]{ \label{fig:word2vec}
		\includegraphics[width=0.25\textwidth]{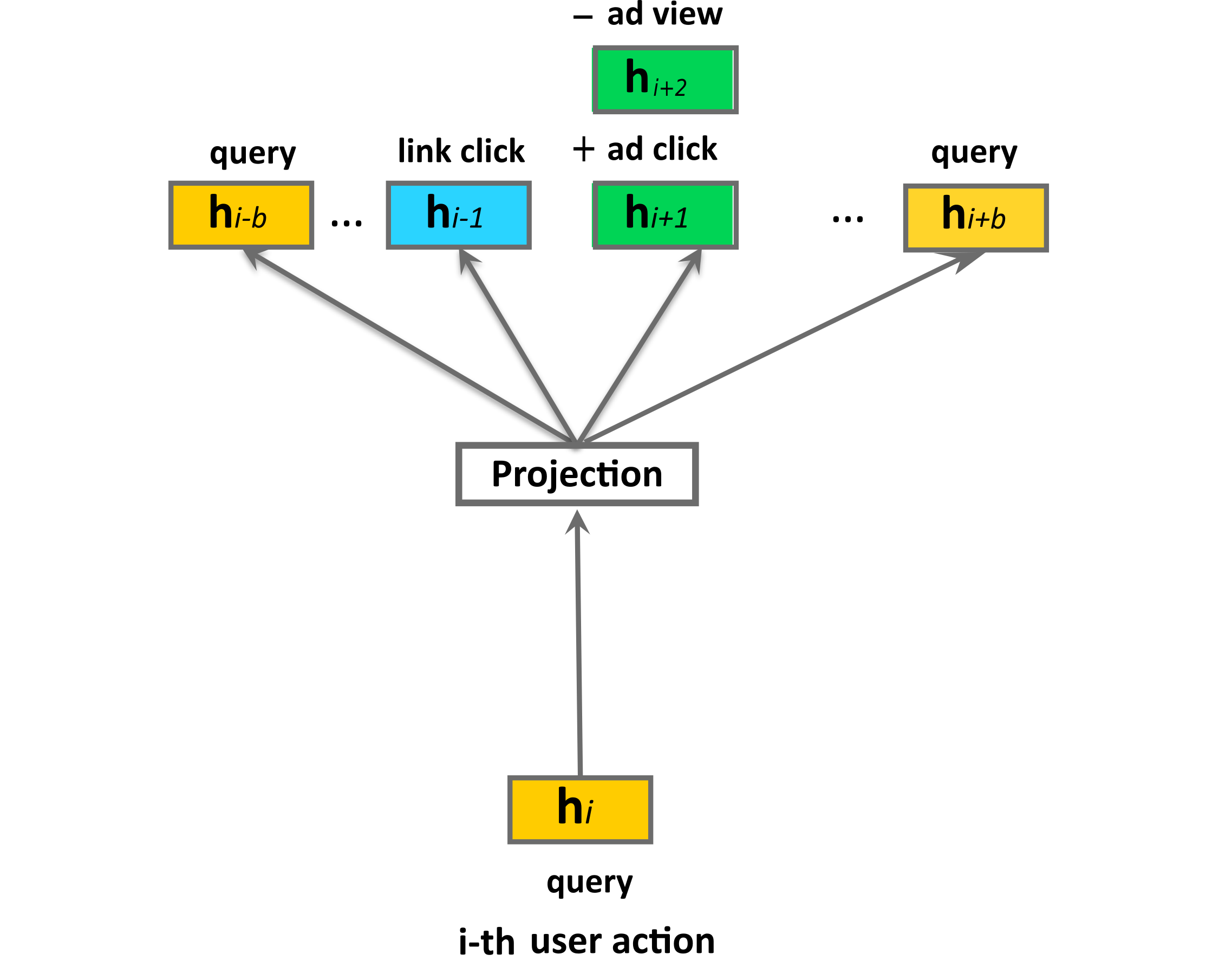}
	}
	\subfloat[{\small location as a global context: gw2v}]{ \label{fig:world2vec}
		\includegraphics[width=0.208\textwidth]{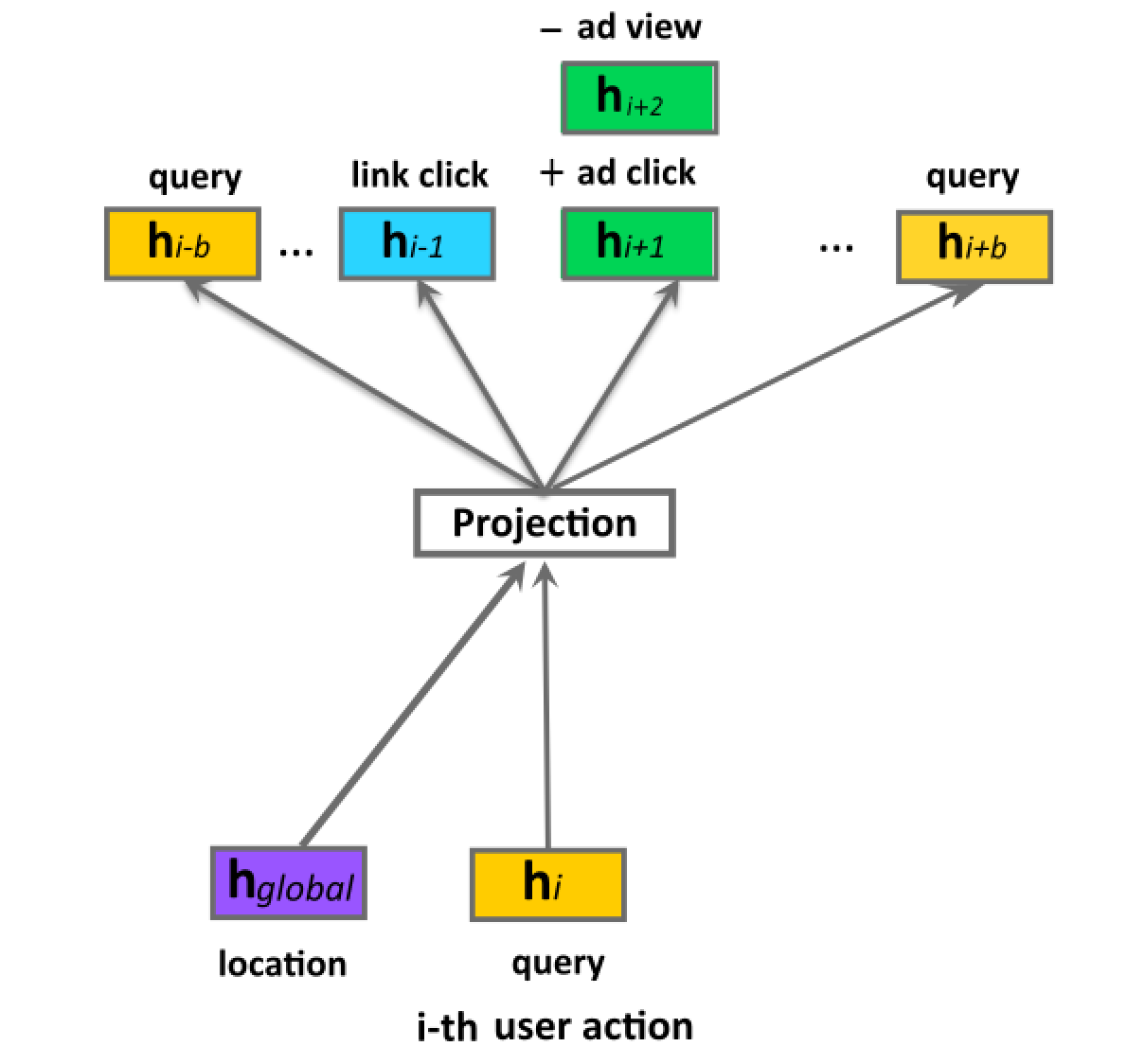}
	}
	\subfloat[{\small location as a local context: lw2v}]{ \label{fig:lworld2vec}
		\includegraphics[width=0.25\textwidth]{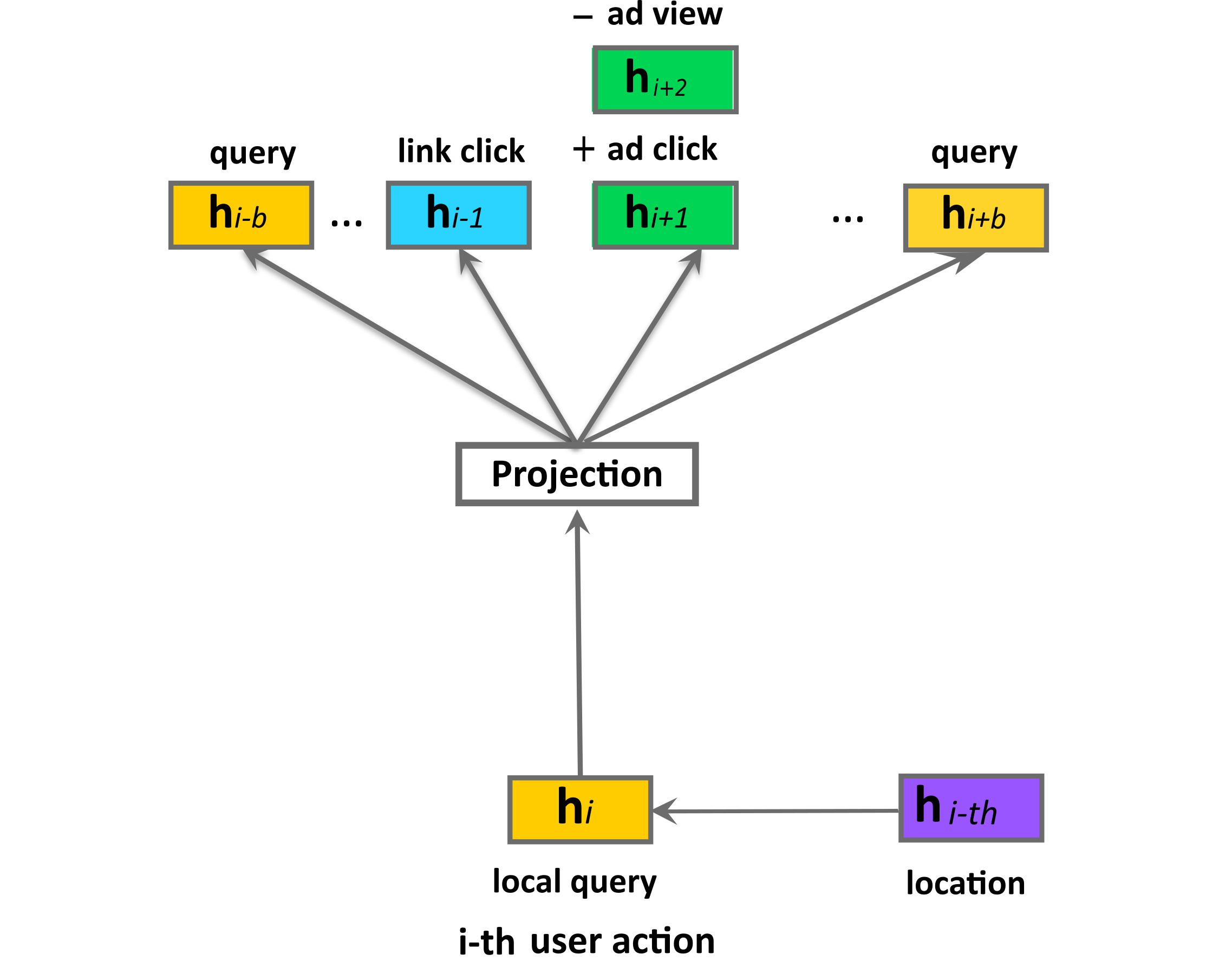}
	}
	\subfloat[{\small location as a part of composition: lw2v+}]{ \label{fig:world2vec_plus}
		\includegraphics[width=0.25\textwidth]{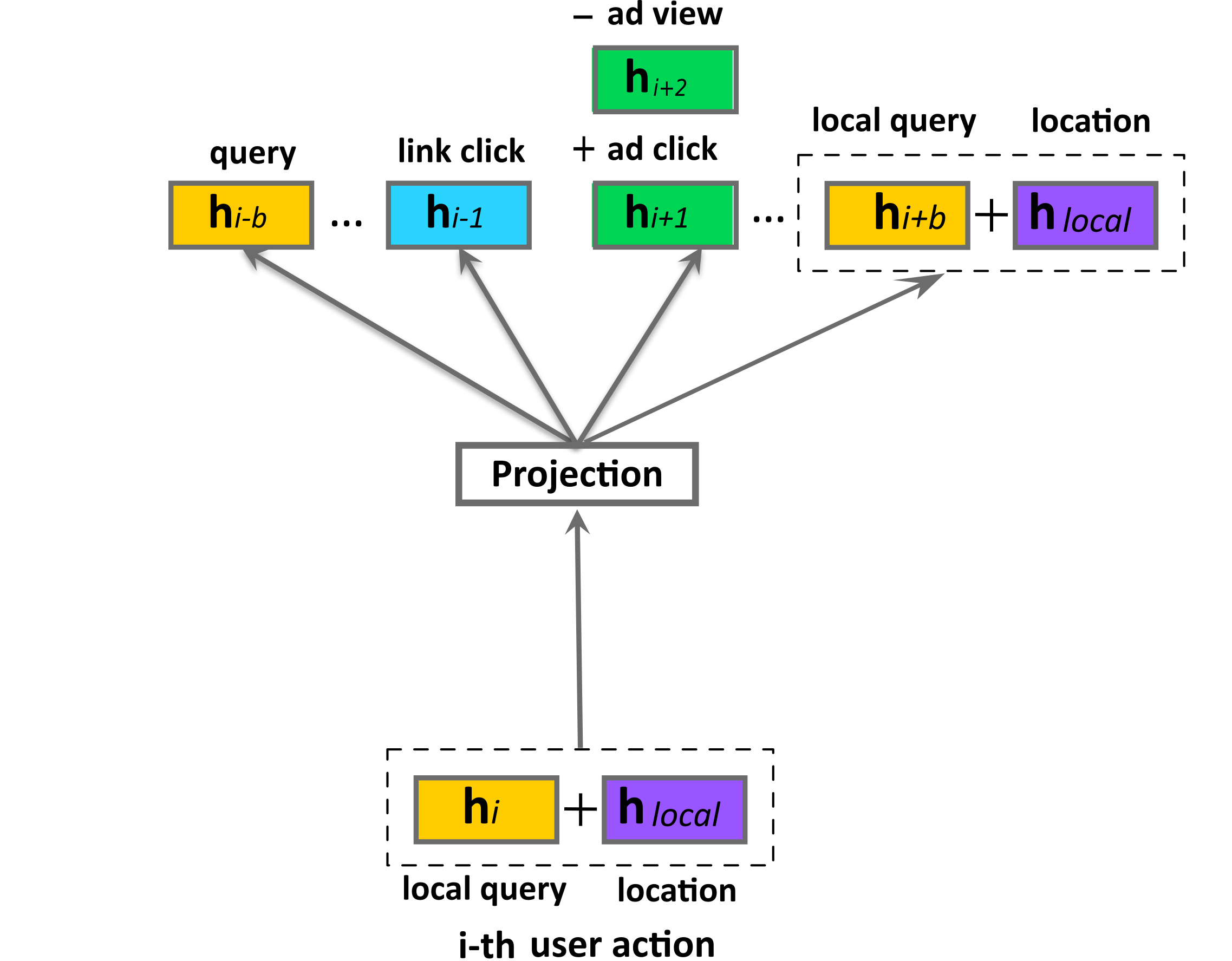}
	}
	\caption{{\small Graphical representations of four models illustrated on a sample session \cite{grbovic2016sigir}.}}
\end{figure*}
In the recent years, there has been an increased focus on learning distributed representations for the unstructured data such as Web search \cite{grbovic2016sigir,ordentlich2016cikm,yin2016}.  Methods such as Latent Dirichlet Allocation (LDA)  have shown benefits over more traditional representations such as TF--IDF bag-of-words (BoW) representation. However, recent advancement~\cite{djuric2015hierarchical,grbovic2016sigir} of Neural Embedding models, based on the popular \textit{word2vec} algorithm~\cite{mikolov2013}, has provided more powerful representations.   
The particular \textit{search2vec} (\textit{s2v}) algorithm \cite{grbovic2016sigir} has repeatedly shown benefits over other state-of-the-art baselines for web search ad retrieval tasks. 
It learns web search tokens: query, search link click (slc) and ad representations from user search session which can be represented as sequences of tokens ordered by their respective time stamp, and observed as a sample from some unknown language. Following this reasoning, the {\it s2v} learns representations of these concepts in a low-dimensional space 
by maximizing the objective function $\mathcal{L}$ over the entire set $\mathcal{S}$ of sessions as follows,
 \begin{equation} \label{word2vec_obj}
 \mathcal{L} = \sum_{s \in \mathcal{S}} \sum_{h_m \in s} \sum_{-b\le i\le b, i\ne 0} \log \P(h_{m+i}|h_m).
 \end{equation} 
The probability $\P(h_{m+i}|h_m)$ of observing a ''neighboring'' token $h_{m+i}$ given the current token $h_m$ is defined using the softmax function over an inner product of $h_{m+i}$ and $h_m$ vectors. 
To allow scalability of the model learning for web search sessions, a negative sampling approach was employed \cite{mikolov2013} and in addition implicit negatives that are observed in the data are used (ad clicks are used as positive samples  and ad views on higher positions with no click  are used as negative samples when updating a query)\cite{grbovic2016sigir}.
The graphical representation of the \textit{s2v} algorithm is given at Figure~\ref{fig:word2vec}. The \textit{s2v} algorithm is used in a particular for the query-ad matching task and it is our main baseline in this study as we build upon the model and results of this method.
However, as discussed above, Web search sessions can be influenced by the user location  or can be focused on the search related to some location. 
Currently, \textit{s2v} is not readily capable of representing location information itself for the query-to-ad matching task, as it sees a whole query as one entity.
In addition, due to heavy tail in local queries distribution, \textit{s2v} as a query history based method cannot address such a query volume.
Therefore, our goal in the following sections is to describe and analyze different approaches of modeling location using distributed representations.

\subsection{Search Session Location as a Global Context: gw2v}
\label{sec:global_models}
To use the \textit{s2v} algorithm for learning distributed representation of location information in addition to query, slcs and ads, we propose a simple and intuitive, but efficient approach to model \textit{user}'s location as a \textit{global context} while learning vectors of other tokens in the Web session. By modeling location as a global context of the Web session, we model dependency between location, query and ads, which ought to improve query-to-ad matching of previous approaches.

\subsubsection{global-worLd2vec model (gw2v)}
The main addition of the \textit{gw2v} model over the \textit{s2v} approach is that location of a user (in the form of either woeid or poi) is modeled as a \textit{global context} of \textit{all} Web search session tokens (therefore, the name global-worLd2vec).
The objective function of the \textit{gw2v} for a given user's location $\textbf{h}_{global}$ for the j-th session is,
\begin{equation} \label{word2vec_obj}
\mathcal{L} = \sum_{s \in \mathcal{S}} \sum_{h_m \in s} \sum_{-b\le i\le b, i\ne 0} \log \P(h_{m+i}|h_m, \textbf{h}_{global}).
\end{equation} 
This way of modeling location naturally surfaces as an intuitive approach for learning location representation for each Web search session and it is graphically represented  in Figure~\ref{fig:world2vec}.

\subsubsection{Experiments with gw2v}
Once the representation of queries, user's location, and ads are learned in a common space, we can find the top K nearest ads to the query vector in that space. For all queries with local intent in the testing month, we retrieve the top K ads and examine if a click happened among them.
We discuss the $query2ad$ results for:
\begin{itemize}[noitemsep,topsep=2pt,parsep=2pt,partopsep=0pt]
\item  $\mathbf{gw2v_{woeid}}$: \textit{gw2v} model with locations represented via woeids,
\item $\mathbf{gw2v_{poi}}$: \textit{gw2v} model with locations represented via pois,
\end{itemize}
in the following paragraphs. \textit{Precision@K} is shown in Table~\ref{tab:gw2v_precisionAtK} for implicit and explicit  queries for $\mathbf{s2v}$,  $\mathbf{gw2v_{woeid}}$ and $\mathbf{gw2v_{poi}}$. In addition to $\mathrm{s2v}$, we compare against 
LDA \cite{Blei2003} (a statistical distributed representation models applied on sessions data to learn representation of queries, ad and search link clicks, used for query-to-ad matching using nearest neighbors search in the topics space)
and TF-IDF \cite{Salton:1986:IMI:576628} (a model that generates broad match candidates by computing the cosine similarity between bag-of-words representation of queries and ads constructed from ad title, description, and display url).
 However, s2v outperformed both of them, with in average
15\% of improvement over TF-IDF and 14\% of improvement over
LDA, so we do not show these results in the table in interest of
space.

\paragraph{$query2ad$ retrieval task results}
We observe that both \textit{gw2v} models outperformed the existing \textit{s2v} model in terms of \textit{precision@K} in all the experiments.
We see, however, that the improvement is more noticeable in the case of implicit local queries ($\sim$6\% up to 20\% of improvement), than on explicit queries ($\sim$2\% up to 6\% of improvement). This makes sense because these models use \textit{user's} location as a global context to all user's actions from the Web session, and in the case of implicit query (such as ``\textit{coffee shops near me}''), this can be helpful. While in the case of explicit queries (such as ``\textit{best hotels in New York City}'', while user's location is Boston), this can actually hurt the algorithm performance, and that is reflected in lower improvement over the \textit{s2v} model.
Even though we do not use location information in the \textit{retrieval} process (we find the closest ad to the \textit{query} vector), the vectors of ads, queries and locations are learned in the common space, thus we see the influence of location representations to relations between queries and ads, resulting in improved performance in $query2ad$ task. 

\begin{table}[h!]
	\centering
	\caption{precision@K for \textit{s2v} and \textit{gw2v} algorithms (for poi and woeid location representations) trained on Web search sessions with local intent (implicit and explicit) queries}
	\label{tab:gw2v_precisionAtK}
	\begin{small}
		\begin{tabular}{ccc|c|c|c|c|c|}
			\cline{4-8}
			& \multicolumn{1}{l}{} & \multicolumn{1}{l|}{} & \multicolumn{5}{c|}{precision@K}           
			\\

			\cline{2-8} 
			\multicolumn{1}{c|}{} & \multicolumn{1}{c|}{model}   & task   & 1      & 2      & 3      & 5      & 10     \\ 
						\hline
			\cline{2-8} 
			\hline
			\multicolumn{1}{|c|}{\multirow{3}{*}{\rotatebox[origin=c]{90}{implicit}}} & \multicolumn{1}{c|}{$s2v$}   & q2ad   
			& 0.204 & 0.369 & 0.490 & 0.550 & 0.582 \\

			\cline{2-8} 
			\cline{2-8} 
						\multicolumn{1}{|c|}{}    & \multicolumn{1}{c|}{{$gw2v_{w}$}} & q2ad &\textit{0.384}&\textit{0.481} &\textit{0.575}& \textit{0.610} & \textit{0.642} \\ 
			\cline{3-8} 

			\cline{2-8}
			\cline{2-8} 
			\multicolumn{1}{|c|}{} & \multicolumn{1}{c|}{{$gw2v_{p}$}}   & q2ad   & \textbf{0.409} & \textbf{0.530} & \textbf{0.628} & \textbf{0.669} & \textbf{0.711} \\ \cline{3-8} 
			\hline
			\hline
			\multicolumn{1}{|c|}{\multirow{3}{*}{\rotatebox[origin=c]{90}{explicit}}} & \multicolumn{1}{c|}{$s2v$}   & q2ad   & 0.062 & 0.075 & 0.086 & 0.094 & 0.111 \\ 
			\cline{2-8}
			\cline{2-8}  
			
						\multicolumn{1}{|c|}{}  & \multicolumn{1}{c|}{{$gw2v_{w}$}} & q2ad   & \textbf{0.111} & \textbf{0.136} & \textbf{0.147} & \textbf{0.158} & \textbf{0.174} \\ 
			\cline{3-8} 	 			
			\cline{2-8} 
			\cline{2-8} 
						\multicolumn{1}{|c|}{}  & \multicolumn{1}{c|}{{$gw2v_{p}$}}   & q2ad   & 0.086 & 0.116 & 0.127 & 0.139 & 0.149 \\ 
			\cline{3-8}  
\hline
		\end{tabular}
	\end{small}
\end{table}

\begin{table}[h!]
	\centering
	\caption{Top 10 nearest search queries to the point of interest ``Universal Studios Florida''}
	\label{tbl:POI_word_cloud}
	{\small
	\begin{tabular}{|l|c|}
		\hline
		 Query                                                 & Cosine Similarity \\ \hline
		 \textit{Universal Islands of Adventure}               & 0.959             \\ \hline
		 \textit{Disney MGM studios Orlando}                   & 0.958             \\ \hline
		 \textit{Universal Studios Orlando discount tickets}   & 0.956             \\ \hline
		 \textit{MGM Studios Orlando}                          & 0.954             \\ \hline
		 \textit{Islands of Adventure Orlando}                 & 0.952             \\ \hline
		 \textit{Harry Potter Universal Studios}               & 0.951             \\ \hline
		 \textit{MGM Studios Orlando theme park}               & 0.950             \\ \hline
		 \textit{Universal Studios Orlando hours of operation} & 0.950             \\ \hline
		 \textit{Disney world epcot attractions}               & 0.949             \\ \hline
		 \textit{Disneyworld Orlando}                          & 0.948             \\ \hline
	\end{tabular}
}
\end{table}

\paragraph{Evaluation of semantic vs. physical location information}
Table~\ref{tab:gw2v_precisionAtK} shows that the overall performance on the $query2ad$ task of the $\mathbf{gw2v_{poi}}$ model is better than that of the $\mathbf{gw2v_{woeid}}$ model. In Table~\ref{tbl:POI_word_cloud}, we show an example of the common vector space of pois, and queries. We retrieved the closest queries to the point of interest: ``\textit{Universal Studios Florida}''. We see queries like: ``\textit{Universal Islands of Adventure}'', ``\textit{Universal Studios Orlando discount tickets}'', ``\textit{mgm studios Orlando}'', ``\textit{Islands of Adventure Orlando}'' etc. The retrieved queries are highly relevant to the given point of interest, which demonstrates that the learned pois representations are meaningful for the $query2ad$ retrieval task.

However, the challenge of working with pois is that their coverage of sessions is lesser than those with woeids ($\sim5\%$, as less users search in the proximity to some Point Of Interest than in a general location for which we can find out physical location in terms of zip, city, etc.). Also, \textit{explicit} local queries in most of the cases will contain locations such as cities (e.g. ``\textit{rent-a-car in San Jose}'') that are better represented via woeids.
For these reasons, we decided to make further improvements in the modeling of locations in local intent queries sessions with woeids instead of pois. 

\begin{figure}[t!]
	\centering
	\includegraphics[width=0.32\textwidth]{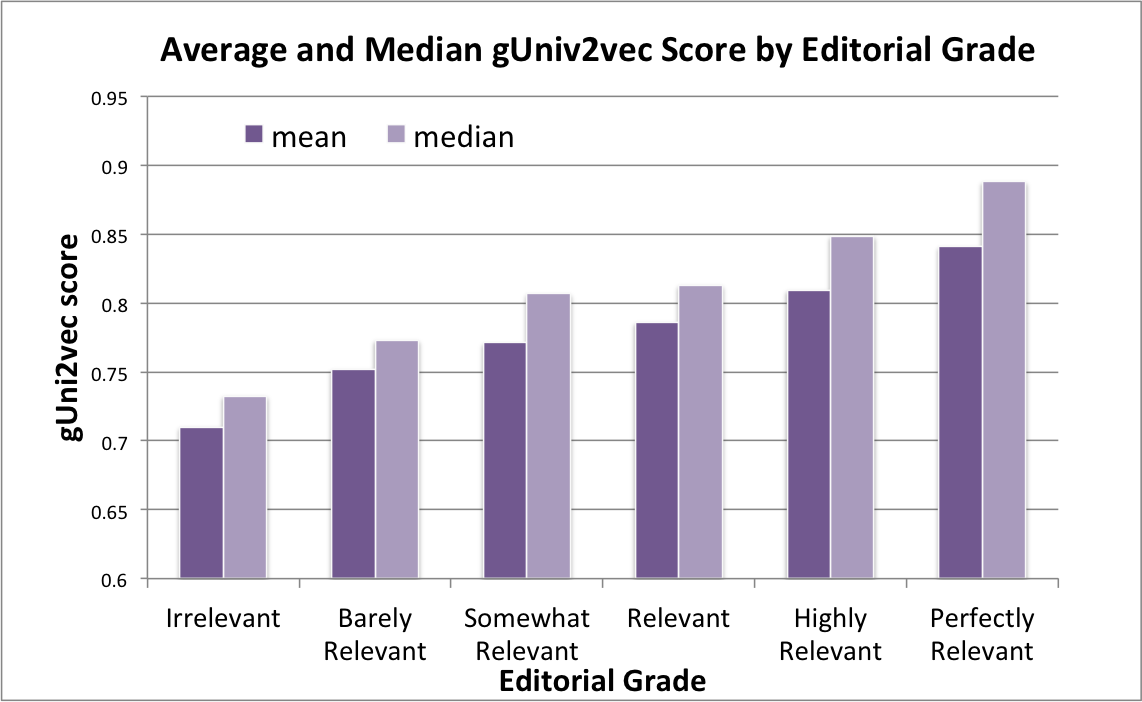}
	\caption{{\footnotesize Average and Median gw2v Score by Editorial Grade}} 
	\label{fig:gUniv2vec_scores}
	\vspace{-5pt}
\end{figure}

\paragraph{Ad ranking quality evaluation}
In addition to the precision@K evaluation for ad retrieval task, we examine query-ad relevance scores of the proposed models. In Figure \ref{fig:gUniv2vec_scores} we show a correlation plot of cosine similarities between vectors of editorially judged queries and ads generated by the \textit{gw2v} method. The cosine similarity scores correlate well with the editorial judgments, indicating a strong potential of the learned representations for the task of retrieving high-quality query-ad pairs.

We further show how different models compare at a particular rank using the average $NDCG$ values over a set of editorial judgments of query-ad pairs. 
\begin{figure}[t!]
	\centering
	\includegraphics[width=0.32\textwidth]{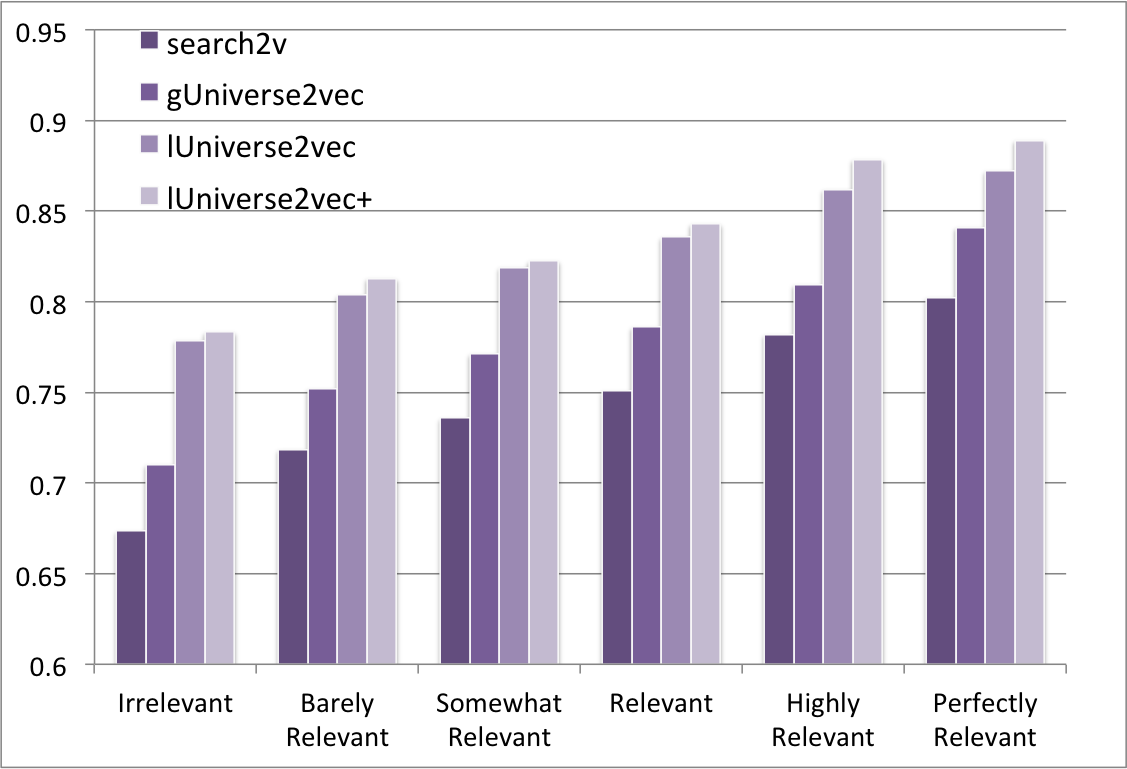}
	\caption{{\footnotesize Average NDCG results for $s2v$, $g2v$, $lw2v$ and $lw2v+$ models}} 
	\label{fig:NDCG_all}
	\vspace{-14pt}
\end{figure}
From Figure~\ref{fig:NDCG_all}, we observe that the \textit{s2v} performs worse than the location aware models (either global or local that are introduced in the following sections). 

\subsubsection{Modeling obstacles}
From the above results, we see that \textit{global} approaches perform well on implicit types of queries, however, they underperformed on explicit types of queries. We can thus conclude that using users' location information for every query in Web search session can be detrimental for some tasks. Therefore, we decide to develop \textit{local} approaches and instead of using \textit{user location} information as a \textit{global context}, use either \textit{query location} for \textit{explicit} queries or \textit{user's phisical location} for \textit{implicit} queries as a \textit{local context} for a local query. 

\subsection{Using Location Information Locally for Local Queries: lw2v}
\label{sec:lw2v_sessions}
Analysis of the previous results show that even though they provide
beneffits in terms of query-ad matching, global models have
shortcomings when it comes to ad retrieval for explicit queries.
In order to address this issue we further model location as a local
context for queries that have local intent.

\subsubsection{lw2v model}
The main difference of the the \textit{lw2v} model compared to \textit{gw2v} is the context modeling. In \textit{lw2v} location information would be used as a \textit{local context} of queries with local intent within the session $s$, and not globally for all elements of the session as in \textit{gw2v}. This way, location vectors $\textbf{h}_{loc}$ and query vector $h_m$ would be learned independently and only when it is given, implicitly or explicitly, that location dictates the part of the session $s$.
The objective function $\mathcal{L}$ of the \textit{lw2v} model can be written as follows,
\begin{equation} \label{word2vec_obj}
\mathcal{L} = \sum_{s \in \mathcal{S}} \sum_{h_m \in s} \sum_{-b\le i\le b, i\ne 0} \log \P(h_{m+i}|h_m,loc_m),
\end{equation} 
where $loc_m$ is defined as
{\footnotesize 	\begin{equation}
	loc_m=\begin{cases}
	\textbf{h}_{loc} & \text{, where \textit{loc} is }
	\begin{cases}
	\text{user location}  & \text{,if $h_m$ is implicit local query} \\
	\text{query location} & \text{,if $h_m$ is explicit local query}
	\end{cases}
	\\
	\textbf{0} & \text{, otherwise}.
	\end{cases}
	\end{equation}
}

The \textit{lw2v} model is shown in Figure~\ref{fig:lworld2vec}, where we clearly see that the local query vector is influenced by the location vector. Such representation allows update of the query vector for relevant locations only, rather than for all queries in the session, thus improving over limitations of explicit queries of the initially intuitive \textit{gw2v} model.

\subsubsection{Experiments with lw2v}
From \textit{lw2v} rows in Table~\ref{tab:local_world2vec_results} we observe improved performance of the \textit{lw2v} model on explicit queries (up to 7\%) over \textit{gw2v} and (up to $\sim$15\%) over \textit{s2v}, showing that using query location instead of user's location can be beneficial for retrieving ads for explicit queries. 
However, that led to degradation in precision for \textit{lw2v} over \textit{gw2v} for implicit queries, still outperforming the \textit{s2v} algorithm (up to $\sim$16\%).
Here, similar to \textit{gw2v}, location information helps in learning better embeddings, in the moment of usage while retrieving ads for specific query. 
In addition to $query2ad$ matching task, we report \textit{precision@K} results for
$(quer\&location)2ad$ task. In order to use location information
in the ad retrieval process, we combine vectors of query and location
and find the closest ad vector to the vector representing their
sum ($\textbf{v}_{query} + \textbf{v}_{location}$) in the $(query\&location)2ad$ task. \textit{Precision@K} results are better on $query2ad$ than on $(quer\&location)2ad$ task, as not training the model for the purpose of retrieval on the composition of vectors affects the performance on this task.

\subsubsection{Modeling obstacles}
Previously described results show that local \textit{lw2v} model does provide improvements for explicit local queries over global \textit{gw2v} model. However, searching for relevant ads based on a local query and location $(query\&location)2ad$ provides no significant improvement over \textit{s2v} model (in most of the cases $\sim$1\% ). To allow the model to learn composable embeddings, whose combinations can be uses to search the vector space, we introduce the model designed for $(query\&location)2ad$ retrieval task.

\cB
\subsection{Learning Compositional Representation of Query and Location: lw2v+}
\label{sec:lw2v+_sessions}
Using location information for local intent queries and with the purpose of testing it as $(query\&location)2ad$, we develop a sum-representation compositional \textit{lw2v} model, the \textit{lw2v+} model. Composition representation in our experiment is defined as sum representation, however it is not limited to this (for reference, please see the overview of compositional embedding models from NLP literature in Section~\ref{sec:word_compositionality} ).
\begin{table}[t!]
	\centering
	\caption{\small{ precision@K on \textit{query2ad}, \textit{(query\&location)2ad},
		\textit{qualifier\_subject\_location2ad}, \textit{subject\_location2ad}, \textit{(subject+location)2ad} and \textit{qualifierType\_subject\_location2ad}  tasks for 3 local-worLd2vec (\textit{l2v},
		\textit{l2v+} and \textit{l2vCRF+}) algorithms trained on Web 
		search sessions with local intent (implicit and explicit) queries or queries semantic fragments} }
	\label{tab:local_world2vec_results}
	{\footnotesize 	\begin{tabular}{lcc|c|c|c|c|c|}
			\cline{4-8}
			& \multicolumn{1}{l}{}                           & \multicolumn{1}{l|}{} & \multicolumn{5}{c|}{\textbf{precision@K}}           \\ \cline{2-8} 
			\multicolumn{1}{l|}{}                            & \multicolumn{1}{c|}{\textbf{model}}                     & \textbf{task}                  & \textbf{1}      & \textbf{2}      & \textbf{3 }     & \textbf{5}      & \textbf{10}     \\ 
			\cline{4-8}
			\hline \cline{4-8}
			\hline
			\multicolumn{1}{|l|}{\multirow{10}{*}{\rotatebox[origin=c]{90}{\textbf{implicit}}}} & \multicolumn{1}{c|}{\multirow{2}{*}{\textbf{{\scriptsize lw2v}}}}     & {\scriptsize q2a}                   & 0.369 & 0.472 & 0.544 & 0.580 & 0.663 \\ \cline{3-8} 
			\multicolumn{1}{|l|}{}                           & \multicolumn{1}{c|}{}                          & {\scriptsize(q$\plus$l)2a}                  & 0.259 & 0.375 & 0.506 & 0.552 & 0.594 \\ \cline{2-8} 
			\multicolumn{1}{|l|}{}                           & \multicolumn{1}{c|}{\multirow{2}{*}{\textbf{{\scriptsize lw2v$\plus$}}}}    & {\scriptsize q2a}                   & 0.375 & 0.485 & 0.553 & 0.593 & 0.656 \\ \cline{3-8} 
			\multicolumn{1}{|l|}{}                           & \multicolumn{1}{c|}{}                          & {\scriptsize(q$\plus$l)2a}                  & \textit{0.388} & \underline{0.492} & \underline{0.573} & \underline{0.620} & \textit{0.672} \\ \cline{2-8} 
			\multicolumn{1}{|l|}{}                           & \multicolumn{1}{c|}		{\multirow{6}{*}{\textbf{{\scriptsize lw2vC$\plus$}}}} 
			& {\scriptsize(q\_s\_l)2a}           & \textbf{0.392} & \textbf{0.497} & \textbf{0.588} & \textbf{0.626} & \textbf{0.689} \\ \cline{3-8} 
			\multicolumn{1}{|l|}{}                           & \multicolumn{1}{c|}{}                          & {\scriptsize(s\_l)2a}              & 0.385 & 0.461 & 0.545 & 0.608 & 0.667 \\ \cline{3-8} 
			\multicolumn{1}{|l|}{}                           & \multicolumn{1}{c|}{}                          &  {\scriptsize(s$\plus$l)2a}               & \underline{0.390} & \textit{0.485} & \textit{0.565} & \underline{0.620} & \underline{0.680} \\ \cline{3-8} 
			\multicolumn{1}{|l|}{}                           & \multicolumn{1}{c|}{}                          &  {\scriptsize(qt\_s\_l)2a}          & 0.384 & 0.446 & 0.562 & 0.611 & 0.672 \\ \cline{3-8} 
			
			\hline
			\hline
			\multicolumn{1}{|c|}{\multirow{10}{*}{\rotatebox[origin=c]{90}{\textbf{explicit}}}} & \multicolumn{1}{c|}{\multirow{2}{*}{\textbf{{\scriptsize lw2v}}}}     & {\scriptsize q2a}                   & 0.124 & 0.157 & 0.177 & 0.201 & 0.226 \\ \cline{3-8} 
			\multicolumn{1}{|c|}{}                           & \multicolumn{1}{c|}{}                          & {\scriptsize(q$\plus$l)2a}                  & 0.096 & 0.134 & 0.159 & 0.185 & 0.217 \\ \cline{2-8} 
			\multicolumn{1}{|c|}{}                           & \multicolumn{1}{c|}{\multirow{2}{*}{\textbf{{\scriptsize lw2v$\plus$}}}}    & {\scriptsize q2a}                   & 0.128 & 0.160 & 0.194 & 0.228 & 0.246 \\ \cline{3-8} 
			\multicolumn{1}{|c|}{}                           & \multicolumn{1}{c|}{}                          & {\scriptsize(q$\plus$l)2a}                  & 0.129 & 0.163 & 0.213 & 0.235 & 0.277 \\ \cline{2-8} 
			\multicolumn{1}{|c|}{}                           & \multicolumn{1}{c|}
			{\multirow{6}{*}{\textbf{{\scriptsize lw2vC$\plus$}}}}
			& {\scriptsize(q\_s\_l)2a}           & 0.116 & 0.168 & 0.227 & 0.262 &\textbf{ 0.297} \\ \cline{3-8} 
			\multicolumn{1}{|c|}{}  & \multicolumn{1}{c|}{} & {\scriptsize(s\_l)2a}              & \underline{0.148} & 0.194 & 0.222 & 0.247 & 0.266 \\ \cline{3-8} 
			\multicolumn{1}{|c|}{}  & \multicolumn{1}{c|}{}                          & {\scriptsize(s$\plus$l)2a} & 0.147 &\textbf{ 0.207 }& \textbf{0.247} &\textbf{ 0.275} & \underline{0.294} \\ \cline{3-8} 
			\multicolumn{1}{|c|}{}                           & \multicolumn{1}{c|}{}                          & {\scriptsize(qt\_s\_l)2a}          & \textbf{0.151} & \underline{0.199} & \underline{0.228} & \underline{0.270} & 0.283 \\ \cline{3-8} 
			\hline
		\end{tabular}}
		\vspace{-10pt}
	\end{table}
	
\subsubsection{lw2v+ model}
The key aspect of the  \textit{lw2v+} model is that conditional probability $\P(h_{m+i}^{(+)}|h_{m}^{(+)})$ is rewritten as combined representation, for example $\P(ad|(query\&location))$ shown in Figure~\ref{fig:world2vec_plus}. Such representation allows for a combined search in the vector space, allowing for more information-rich queries. 
		
WorLd2vec models learn vector representation for each token in Web search session. To account for dependency of the tokens in the \textit{lw2v+} model, we propose compositional tokens $h^{(+)}$ to be represented as a set $\mathcal{G}_h = \{t_1, \ldots, t_G\}$ for $G$ tokens provided in the composition.
The dot product $\mathbf{v}_{h_m}^\top \mathbf{v}_{h_{m+i}}^\prime$ used in Softmax for modeling $\P(h_{m+i}|h_m)$, now becomes: $\sum_{t_g \in \mathcal{G}_h} \textbf{v}_{t_g}^\top \mathbf{v}_{h_{m+i}}^\prime$ (for the example in which composition token is in the context). The softmax function is in that case:
\begin{equation}\label{eq:softmax_plus}
\P(h_{m+i}|h_m^{(+)}) = \frac{\exp(\sum_{g \in \mathcal{G}_h} \textbf{v}_g^{T}\mathbf{v}_{h_{m+i}}^\prime)}{\sum_{d=1}^{|\mathcal{D}|} \exp(\sum_{g \in \mathcal{G}_h} \textbf{v}_g^{T} \mathbf{v}_{d}^\prime)}.
\end{equation}
In equations above, composition token was a context token, but it can be a central token instead, or in general case, both context and central tokens can be composition tokens.
	
The objective function $\mathcal{L}$ of the \textit{lw2v+} model can be written as follows,
\begin{equation} \label{lw2v+_objective}
\mathcal{L} = \sum_{s \in \mathcal{S}} \sum_{h_m \in s} \sum_{-b\le i\le b, i\ne 0} \log \P(h_{m+i}^{(+)} | h_{m}^{(+)}).
\end{equation} 
where $h_{m}^{(+)}$  is defined as
\begin{equation}
	h_m^{(+)}=\begin{cases}
	(h_{query}+h_{loc})_m & \text{, if \textit{query} is a local intent query } \\ & \text{and \textit{loc} is it's location}\\
	h_{m} & \text{, otherwise}
	\end{cases}
\end{equation}

\subsubsection{Experiments with  lw2v+}
	Table~\ref{tab:local_world2vec_results} shows \textit{precision@K} results of the \textit{lw2v+} model for \textit{query2ad} and $(query\&location)2ad$ tasks, respectively. 
	We can now see the improvement in precision for $(query\&location)2ad$ task over \textit{query2ad} 
	retrieval task for this model, showing that not only the model was able to capture useful location information while learning the vector embeddings, but was able to exploit this information in an efficient manner for the retrieval task as well.
	We also see that this way of learning vector representations helped in both tasks when compared to the \textit{lw2v} model, 
	especially on  $(query\&location)2ad$ task, 
	as the improvement over \textit{lw2v} is up to 13\%.
	Improvement over \textit{s2v} ranges from $\sim$5\% up to  $\sim$20\%, 
	while for the \textit{gw2v} model results are comparable for implicit queries on \textit{query2ad} task, but in most of the cases \textit{l2v+} on the $(query\&location)2ad$ task performs better than
	\textit{gw2v} on the \textit{query2ad} task. On explicit queries this improvement is certain and it goes up to 16\% on both tasks.

\begin{figure}[t!]
	\centering
	\subfloat[lw2v]{ \label{fig:sfa_local}
		\includegraphics[width=0.33\textwidth]{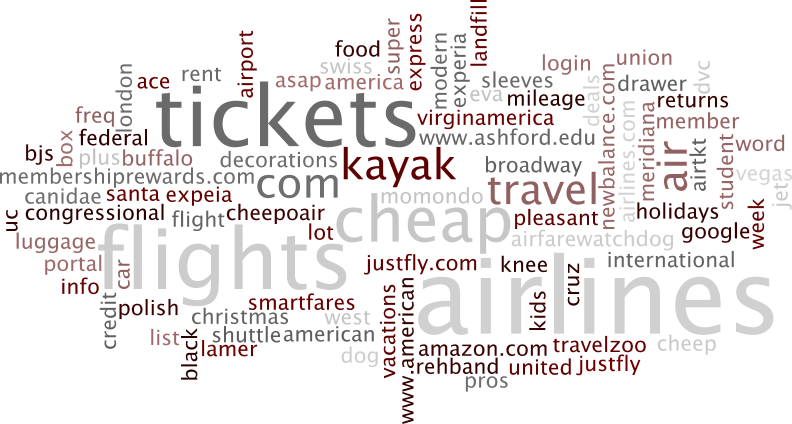}
	}	\vspace{-11pt}
	\subfloat[lw2v+]{ \label{fig:atl_local_plus}
		\includegraphics[width=0.33\textwidth]{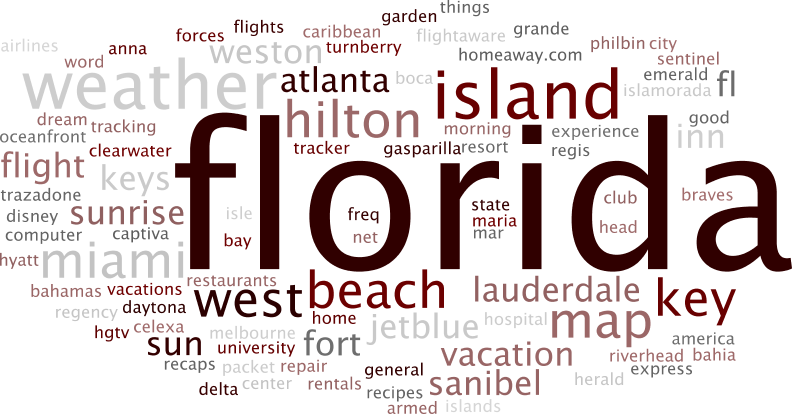}
	}
	\caption{{\footnotesize Individual keywords for top 100 relevant queries to the search of ATL airport, obtained by \textit{lw2v} (a) and \textit{lw2v+} (b)}}
	\label{fig:keywords_airports}
	\vspace{-19pt}
\end{figure}

\paragraph{Location relevant keywords}
Further, we analyze learned vector representations of query and location tokens. We focus our use case on the Atlanta airport (ATL) woeid (as the largest airport in the world) and analyze proximity of queries in the embedded space to this location vector given by \textit{lw2v} and \textit{lw2v+} models. Due to space limitation, we constrain the analysis to one example, but we observe similar traits on other airport examples as well. 
We compare embeddings obtained by these two models by retrieving top 100 nearest queries in the embedded space and provide word-clouds of individual words found in the top 100 queries for each model in Figure~\ref{fig:keywords_airports}.
Figure~\ref{fig:sfa_local} shows query words relevant for the local search of the ATL airport. We can see that \textit{lw2v} model was capable of learning correlation of airport embeddings with queries that primarily focus on purchasing airport tickets and obtaining information on flights (we observe high frequency of words such as 'airlines', 'flights' and 'tickets'). 
This, however intuitive, does not entirely reflect the actual intent users may have while searching from airport (implicit local queries) or about airport (explicit local queries). 
Queries made while users were physically near the airport are not likely made with intent of purchasing a ticket. Rather, users make both explicit and implicit local queries to inform themselves about desired destinations. These destinations are seemingly geographically dependent, which is exactly shown in word-clouds of the \textit{lw2v+} model (Figure \ref{fig:atl_local_plus}). 
\textit{lw2v+} model learned high relevance of popular destinations travelers traveling from the airport are likely to visit,
like 'Florida', 'Miami', 'Key West', 'Bahamas', 'Atlanta', or hotels, resorts, transportations from the airport to a destination, like 'inn', 'resort' and names of particular hotels. Of course, queries related to travel, like 'flight' and 'airport' are in the closest proximity, confirming that plus models is also capable of retrieving flight ads as the \textit{lw2v} model, with the enriched context of user's local intent.
From this analysis we can conclude that \textit{lw2v+} model is capable of learning embeddings more relevant to users local travel intents compared to the \textit{lw2v} approach.
	
\subsubsection{Modeling obstacles}
Even though this model outperforms previous approaches overall, it is still susceptible to large variability in query writing and sensitive to cold start problems. Therefore, we further improve \textit{worLd2vec} family of models by reducing variability and handling novel queries.
	
\subsection{Learning Compositional Representations of  Query Extractions and Location: lw2vCRF+}
\label{sec:lw2v_crf+_sessions}
In order to achieve this goal, we parse the local queries to associate semantic tags for query fragments.
This task is often referred to as information extraction from short text or shallow semantic parsing. Conditional Random Fields \cite{lafferty2001conditional} (CRF) is one of the best performing  algorithms for extracting information from short text \cite{li2009extracting}.
We use 28,000 random local queries that are annotated by human experts and labeled according to the BIO scheme for the following tags: Organization Name, Business Category and Location (state, city and zip levels).  For extracting qualifiers and attributes, we rely on the longest string matched on a hand-curated dictionary of popular qualifiers and attributes. The performance of the CRF model for the tags avove is noted in Table~\ref{tbl:crfperf}.
	\begin{table} [h!]
		\centering
		\caption{Performance of the local CRF model}
		\vspace{-10pt}
		{\small 	\begin{tabular}{|c|c|c|}
				\hline
				Tag & Precision & Recall   \\ \hline
				Organization Name &  0.89&   0.86  \\ \hline
				Business Category &  0.7 & 0.74  \\ \hline
				Location:State&  0.96 &  0.85   \\ \hline
				Location:City &  0.92 &    0.93 \\ \hline
				Location:ZipCode & 1.0 & 0.97 \\ \hline 
		\end{tabular}}
		\label{tbl:crfperf}
		\vspace{-11pt}
	\end{table}
We then annotate \textit{Organization Name} and  \textit{Business Category} tags as \textit{subject}s, \textit{Location} tag (if less than a state level: e.g city, zip, neighborhood, etc) as \textit{location}, Qualifier tags (except location related qualifier, e.g. 'near me') as \textit{qualifier}s and use their type to have another \textit{qualifierType} annotation. We further parse queries using those annotations and form the following web session tokens out of the semantic query fragments ordered by matching priority: 
	\begin{enumerate} [noitemsep,topsep=2pt,parsep=2pt,partopsep=0pt]
		\item qualifier\_subject\_location (i.e. best\_hotels\_woeid\_2459115)
		\item subject\_location (i.e. hotels\_woeid\_2459115)
		\item subject + location  (i.e. hotels + woeid\_2459115)
		\item qualifierType\_subject\_location \\(i.e. superlative\_hotels\_woeid\_2459115)
	\end{enumerate}
	If a query was never observed before, the extracted tokens ordered by matching priority are used to find the most relevant ads, thus solving the cold-start problem.
	
	In order to learn semantic query fragments properly, 
	our goal is now to minimize the distance between the vector of an explicit local query and the summation of vectors of implicit version of that query and location. In example of explicit query ``\textit{best hotels in New York}'' and implicit version of that query ``\textit{best hotels near me}'' searched from \textit{New York}, we would like for the vector of the explicit query $\mathbf{{v}_{q\_e}}$, i.e. (best\_hotels\_in\_new\_york) and summation of the vectors of implicit query $\mathbf{{v}_{q\_i}}$ and location $\mathbf{{v}_{loc}}$, i.e. ($\mathbf{v}$(best\_hotels\_near\_me) + $\mathbf{v}$(woeid\_2459115)) to be close in the embedded space:
	\begin{equation}
	\label{eq:minD}
	min D( \mathbf{{v}_{q\_i}} + \mathbf{{v}_{loc}} , \mathbf{{v}_{q\_e}})
	\end{equation}
	
	To map the representations of query fragment tokens close in the embedded space, they should be exposed to similar context, thus we propose a new version of the \textit{worLd2vec} objective function.
	
\subsubsection{lw2vCRF+ model}
	
	We define the conditional probability $\P(h_{m+i}^{(c+)}|h_{m}^{(c+)})$ in the \textit{lw2vCRF+} model to be calculated from combined representation of set $\mathcal{R}$ of extraction tokens $e_{m, r}$ from local query semantic fragments.
	The objective function $\mathcal{L}$ of the \textit{lw2vCRF+} model can be written as follows,
	
	\begin{equation} \label{lw2vcrf+_objective}
	\mathcal{L} = \sum_{s \in \mathcal{S}} \sum_{h_m \in s} \sum_{-b\le i\le b, i\ne 0} \sum_{r =1,...,R_m} \log \P(h_{m+i}^{(c+)}|e_{m, r}^{(c+)}),
	\end{equation} 
	where $\P(h_{m+i}^{(c+)}|e_{m, r})$ represents probability
	of observing a session event
	$h_{m+i}^{(c+)}$ given the $r$-th extraction combination from local query $e_{m, r}^{(c+)}$. To generalize, $h_m^{(c+)}$ is defined as:
	\begin{equation}
	h_m^{(c+)} = \begin{cases}
	\left\lbrace e_{m,r}\right\rbrace \in \mathcal{R} & \textnormal{, if } query \textnormal{ is local query and } \left\lbrace e_{m,r}\right\rbrace \\ 
	&  \textnormal{ is a set of it's extraction tokens} \\ 
	\\
	h_{m} & \textnormal{, otherwise}
	\end{cases}
	\end{equation}
	and $\P(h_{m+i}^{(c+)}|e_{m, r})$ is calculated as a product of probabilities
	$\P(h_{m+i}^{(c+)}|e_{m, r}) =  \P(e_{m+i,1}|e_{m, r}) \times ... \times \P(e_{m+i,R_{m+i}}|e_{m, r})$ each defined using (\ref{eq:softmax_plus}).
	As a result, tokens for semantic query fragments, location and ads that often co-occur and have similar contexts will have similar representations as learned by this model.
	\begin{figure}[t!]
		\centering
		\includegraphics[width=0.35\textwidth]{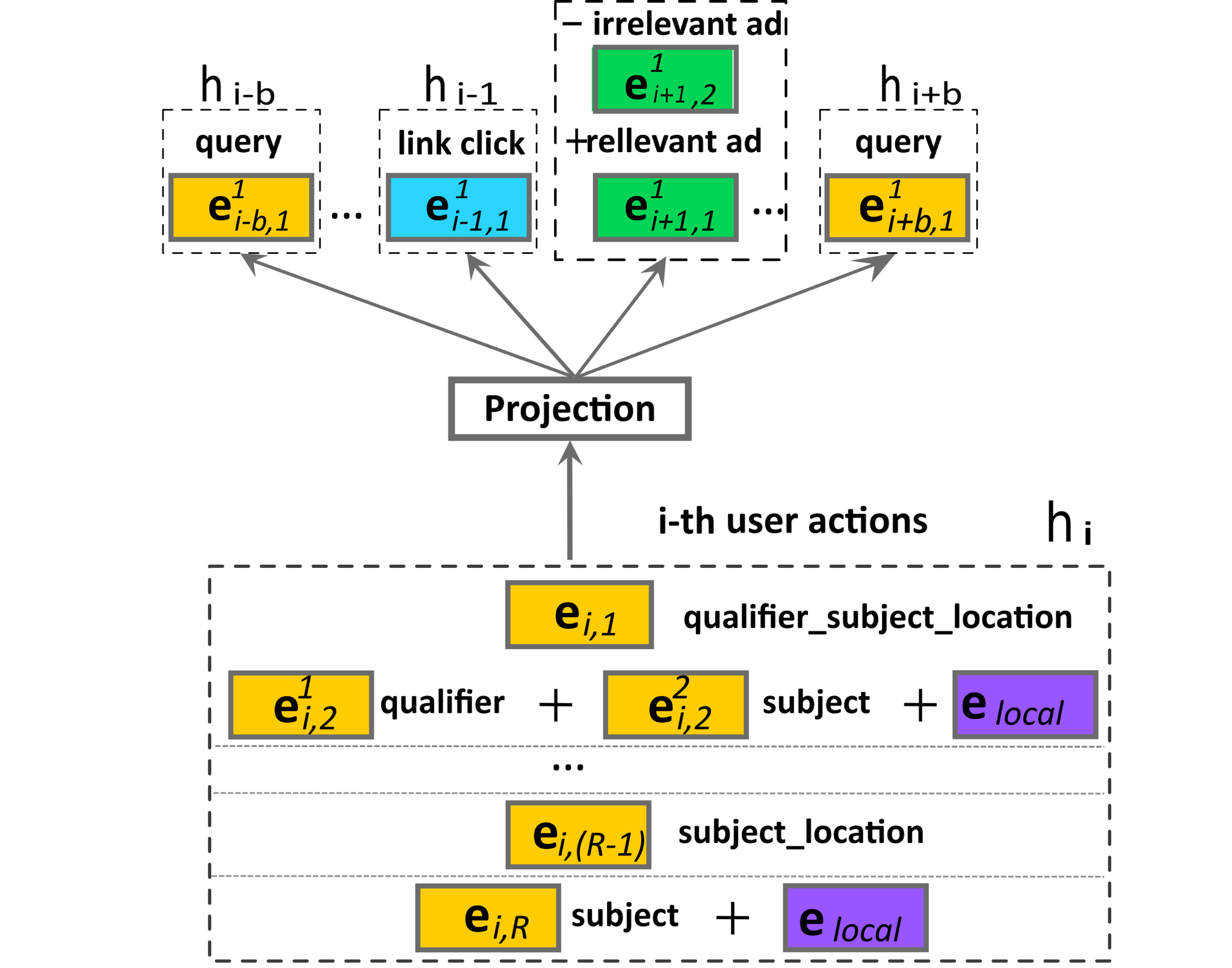}
		\vspace{3pt}
		\caption{{\footnotesize Graphical representations of the \textit{lw2vCRF+} model illustrated on a sample session. The representations of semantic query fragments are learned as a sum composition.}} 
		\label{fig:world2vec_CRFplus}
		\vspace{-15pt}
	\end{figure}
	
\subsubsection{Experiments with lw2vCRF+ model}
	Table~\ref{tab:local_world2vec_results}  \textit{lw2vCRF+} rows show \textit{Precision@K} for this model. 
	We can draw several conclusions from them:
		\begin{enumerate}
	\item when compared to the \textit{lw2v+} model and its $(query\&location)2ad$ task, especially in the case of explicit queries,  \textit{(subject+location)2ad} of the \textit{lw2vCRF+} model retrieves more accurate predictions for ads, 
	demonstrating its ability to cope with the noise from query writing presentation by focusing to query's key elements; 		
	\item As in the case of \textit{lw2v+} (on \textit{query2ad} vs. $(query\&location)2ad$ tasks), if the similar amount of information is used, in this case subject and location, compositional representation gives more accurate results than simple fragment concatenation, i.e. \textit{subject + location} vs. \textit{subject\_location} representation gives up to 3\% improvements in precision;
	\item If additional information is available to match the query, for instance qualifier\_subject\_location, for implicit queries it leads to the best performance in ad retrieval task. Coverage is, however, less ($\sim6.23\%$ of all local queries) but if the information about qualifier is present for implicit queries, it should be used as it retrieves the most relevant ads.
		\end{enumerate}
	
	To explore the vector space and validate that the model was able to embed compositional representations close to their concatenation pair as in Eq.\ref{eq:minD},  we retrieve the closest semantic query fragments to the summation vector of subject \textit{'hotels'} and location \textit{'woeid\_2459115'} (New York): 
	\begin{table}[h!]
		\centering
		\vspace{-10pt}
		\caption{The closest semantic query fragments to the summation
vector of subject '\textit{hotels}' and location '\textit{woeid\_2459115}' (\textit{New York})}
		\label{fig:hotels_nyc}
		\small{
		\begin{tabular}{|l|c|}
			\hline
			\textbf{Query Composition}                  & \textbf{Cosine Similarity} \\ \hline
			\textit{woeid\_249115}                      & 0.953                      \\ \hline
			\textit{hotels\_woeid\_249115}              & 0.947                      \\ \hline
			\textit{hotel\_woeid\_249115}               & 0.921                      \\ \hline
			\textit{superlative\_hotels\_woeid\_249115} & 0.906                      \\ \hline
			\textit{hotel\_woeid\_12589342}             & 0.901                      \\ \hline
		\end{tabular}
	}
	\end{table}

	We observe that the first retrieved fragment is actually the location itself, and that the second is the \textit{subject\_location} concatenation pair, while the rest is a variation (the location within New York, like Manhattan; concatenation with the superlative or singular form of the central fragment). In average cosine similarity score between concatenation and comopsitional representations is 0.911, demonstrating the ability of the method to learn meaningful compositional representations of tokens in the common space of semantic query fragments, locations and ads.

\section{Related Work}
\label{sec:related+word}
Here we introduce related work in three different units: retrieval and ranking of local queries, location representations,  and recent advances in neural language models and compositional distributed embeddings.

\subsection{Retrieval and Ranking for Local Queries} 
Related work in retrieval and ranking for local queries can be grouped further into two related areas: (i)  geographic  information retrieval,  and (ii) inference from location Web session data from various sources, such as  page content and  search engine logs. 

i) Geographic information retrieval tasks key challenge is how to retrieve geographic information correctly and efficiently in location-based search services.
In ~\cite{Chen2006}, the authors propose query processing techniques for geographic information retrieval by employing efficient data structures and algorithms. Further, in ~\cite{Yi2009}, the authors propose an approach for automatic binary classification of local intent queries at the fine-grained geographic level of city location. We use this approach in our study to detect local intent queries. 
ii) A second group of related tasks focuses on using location information to improve Web services.
For location-aware document retrieval in search ~\cite{Bennett2011} and \cite{li2006indexing} explore how the user's location affects document relevance while investigating how to enrich location representation. Direction based queries (directions from a to b) especially benefit from using location information, and ~\cite{Venetis2011} proposed approach for modeling location popularity using these queries by taking into account distance between a user and a place.
iii) Finally, for computational online advertising, ~\cite{Lv2012} investigate how the physical distance of users to a place of interest, number of reviews, and rating, affect the click-through rate.

\subsection{Location Representation} 
In this section, we discuss related research that focuses on richer representation of user location and techniques to exploit the representation for retrieval and ranking tasks.  

Distance between the user's location and the business location has always been considered as an important feature for Geo IR~\cite{Yi2009}.
Berberich et al~\cite{Berberich2011}  propose that in addition to the distance between user and business, category of business being ranked, logs of driving direction requests, customer ratings and logs of accesses to business web sites for popularity can be used to improve ranking popularity of a local business.
An et al~\cite{An2016} also address the location representation, although for a different use case (cold start), using geographic features of businesses to predict the relevance of a business to a user query.

Associating semantic meaning to the user's current physical location, like home, restaurant or store, is of interest to location aware services like  FourSquare (check in experience), Instagram (auto tag photos with location name) and so on.  Shaw et al ~\cite{Shaw2013} address the problem of mapping a noisy estimate of the user's physical location to a semantically meaningful point of interest by modeling user features and location features. Similarly to~\cite{An2016}  and  ~\cite{Berberich2011},  the relationship between the user and business location is modeled using features richer than Euclidean, Manhattan or Haversine distance. Spatial distribution of historical check-ins  and temporal checkin activity are presented as key features to model a venue. 

\subsection{Learning Vector Composition Models}
\label{sec:word_compositionality}
In this section, we discuss the background research related to richer word and phrase representation via exploring composition of word/subword vectors.
These approaches show promising results in vector composition which we use to motivate our approaches for compositional query and location to ad ($query\&location2ad$) retrieval inference task. 

In ~\cite{LebretC15} the authors propose a model that jointly learns word vector representations and their summation with autoencoders. To embed phrases with different sizes into a common semantic space, the average of word vector representations is found and the aim is to learn to discriminate whether words are in the phrase chunk or not.
In  ~\cite{BojanowskiGJM16}, the authors propose an approach to represent morphologically rich language words as a sum of morphemes for word analogy tasks.
In ~\cite{yu2015learning}, phrase embeddings are constructed by learning how to compose word embedding using features that capture phrase structure and context. Here, the phrase composition is a weighted summation of component words' emebeddings. 
Finally, ~\cite{PengG16} introduced a general framework for joint learning of distributed word vector representations via skip-gram model and their way of composing to form phrase embeddings, where compositional functions can be linear combination,  summation,  concatenation etc.

\section{Conclusions}
\label{sec:conclusion}
We proposed a family of methods based on neural language models to leverage available user and/or query location information and learn low-dimensional representations of search queries, semantic query fragments, locations and ads based on contextual co-occurrence in user search sessions. 
When compared to the baseline approach, we showed that the proposed worLd2vec models generated more relevant queries and location to ads matches, and had higher precision, suggesting higher monetization potential for sponsored search for queries with local intent than of the baseline. 

\bibliographystyle{abbrv}
\bibliography{worLd2vec} 
\balance

\end{document}